\documentclass[12pt, centerh1]{article} % onecolumn (second format)

\usepackage{graphicx,colonequals}
\usepackage{natbib}
\usepackage{amsmath, amsthm, amssymb, amsfonts}
\usepackage{supertabular}
\usepackage{longtable, multirow}
\usepackage{rotating}
\usepackage{subfigure}
\usepackage{epsfig}
\usepackage{color}

\newcommand{\tr}{\,\mbox{tr}}
\newcommand{\load}{\mathbf\Lambda}
\newcommand{\noisev}{\mathbf\Psi}

\newcommand{\vecx}{\mathbf{x}}

\newcommand{\veck}{\mathbf{k}}
\newcommand{\vecv}{\mathbf{v}}
\newcommand{\vecq}{\mathbf{q}}
\newcommand{\vecX}{\mathbf{X}}

\newcommand{\vecd}{\mathbf{d}}
\newcommand{\veca}{\mathbf{a}}

\newcommand{\vecV}{\mathbf{V}}
\newcommand{\vecD}{\mathbf{D}}
\newcommand{\vecY}{\mathbf{Y}}
\newcommand{\vecC}{\mathbf{C}}

\newcommand{\vecA}{\mathbf{A}}
\newcommand{\vecU}{\mathbf{U}}

\newcommand{\vecs}{\mathbf{s}}
\newcommand{\vecr}{\mathbf{r}}

\newcommand{\vecB}{\mathbf{B}}

\newcommand{\vecy}{\mathbf{y}}

\newcommand{\vecu}{\mathbf{u}}

\newcommand{\vecmu}{\mbox{\boldmath$\mu$}}
\newcommand{\vecLambda}{\mathbf{\Lambda}}
\newcommand{\vecPsi}{\mathbf{\Psi}}
\newcommand{\vecPhi}{\mbox{\boldmath$\Phi$}}

\newcommand{\vecDelta}{\mathbf{\Delta}}
\newcommand{\vecOmega}{\mathbf{\Omega}}
\newcommand{\vectheta}{\mbox{\boldmath$\theta$}}

\newcommand{\vecepsilon}{\mbox{\boldmath$\epsilon$}}
\newcommand{\vecdelta}{\mbox{\boldmath$\delta$}}
\newcommand{\vecalpha}{\mbox{\boldmath$\alpha$}}
\newcommand{\vecvarthet}{\mbox{\boldmath$\vartheta$}}

\newcommand{\matsig}{\mathbf{\Sigma}}

\title{Mixtures of Hidden Truncation\\ Hyperbolic Factor Analyzers}  
  \author{Paula M.\ Murray$^{*}$, Ryan P.\ Browne$^{**}$ and Paul D.\ McNicholas$^{*}$}
  \date{\small $^{*}$Department of Mathematics and Statistics, McMaster University, Ontario, Canada.\\
 $^{**}$Department of Statistics and Actuarial Sciences, University of Waterloo, Ontario, Canada.}
 
\begin{document}

\maketitle 

\begin{abstract}
The mixture of factor analyzers model was first introduced over 20 years ago and, in the meantime, has been extended to several non-Gaussian analogues. In general, these analogues account for situations with heavy tailed and/or skewed clusters. An approach is introduced that unifies many of these approaches into one very general model: the mixture of hidden truncation hyperbolic factor analyzers (MHTHFA) model. In the process of doing this, a hidden truncation hyperbolic factor analysis model is also introduced. The MHTHFA model is illustrated for clustering as well as semi-supervised classification using two real datasets.\\

\noindent \textbf{Keywords}: hidden truncation hyperbolic distribution; hidden truncation hyperbolic factor analysis; MHTHFA; mixture of factor analyzers; mixture of hidden truncation hyperbolic factor analyzers.
\end{abstract}

\section{Introduction}

Model-based clustering is an effective tool for identifying homogeneous subpopulations within a heterogeneous population.  Although often employed for cluster analysis, mixture modelling approaches are traditionally ill-suited to modelling high-dimensional data sets due to the prohibitively large number of model parameters that must be estimated. Given that modern technology allows us to collect and store vast amounts of data with ease, mixture models must be adapted to handle high-dimensional data. 
The mixture of factor analyzers (MFA) model \citep{ghahramani97,mclachlan00a} reduces the number of parameters to be estimated by introducing latent factors. A number of other models have been developed based on the MFA model and place additional restrictions on the component covariance parameters. These include the mixture of probabilistic principal component analyzers model \citep{tipping99b} and the family of parsimonious Gaussian mixture models (PGMM) of \cite{mcnicholas08,mcnicholas10d}. Note that all of the aforementioned models are developed based on Gaussian mixtures. These approaches, as well as some others, are covered within the excellent review of work on model-based clustering of high-dimensional data given by \cite{bouveyron14}. 

Some work has been carried out extending the MFA to $t$-mixtures \citep[e.g,][]{peel00,andrews11a,andrews11c,steane12,lin14} and, in recent years, there has been a surge in interest in using non-Gaussian distributions to develop mixture models capable of detecting asymmetric clusters. This includes the mixture of shifted asymmetric Laplace (SAL) distributions \citep{franczak14}, mixture of skew-normal distributions \citep{lin09}, mixture of skew-$t$ distributions \citep{vrbik12,vrbik14,lee14}, the mixture of normal inverse Gaussian (NIG) distributions \citep{karlis09,subedi14}, mixture of variance-gamma distributions \citep{smcnicholas17}, mixture of generalized hyperbolic distributions \citep{browne15}, mixture of joint generalized hyperbolic distributions \citep{tang18}, and a mixture of coalesced generalized hyperbolic distributions \citep{tortora19} amongst others. A recent review of work in model-based clustering, including some coverage of non-Gaussian mixture models that have appeared in the literature to date, is given by \cite{mcnicholas16b}. 

When one considers the various formulations of skewed distributions that have appeared in the model-based clustering literature, it is notable that there are different ways of imposing skewness within the component densities. For example, skewness can be introduced based on a mean-variance normal mixture or via hidden truncation. \cite{franczak14} use a scale mixture of normals to develop a mixture of SAL distributions. \cite{murray14b,murray14a} use a mean-variance mixture of normal distributions to develop a mixture of skew-$t$ distributions, \cite{browne15} follow a similar approach for a mixture of generalized hyperbolic distributions, as do \cite{karlis09} for a mixture of NIG distributions and \cite{smcnicholas17} for a mixture of variance-gamma distributions. \cite{lin09} and \cite{lin10} develop a multivariate skew-normal mixture model based on the truncated-normal distribution and a multivariate skew-$t$ mixture model based on the truncated $t$-distribution, respectively. \cite{lee16} use a mixture of canonical fundamental skew-$t$ (CFUST) distributions \citep[see][for details on the CFUST distribution]{arellano05}. %\cite{murray17b} introduce the hidden truncation hyperbolic (HTH) distribution, which contains the CFUST and many other distributions as special or limiting cases. 
\cite{murray17b} introduce the hidden truncation hyperbolic (HTH) distribution. The HTH approach is based on a truncated hyperbolic random variable, and \cite{murray17b} demonstrate the effectiveness of a mixture of HTH distributions for clustering.
The HTH distribution contains certain formulations of skew-$t$ and skew-normal distributions as limiting cases. In particular, the HTH distribution includes the CFUST distribution as a limiting case and, as a consequence, the canonical fundamental skew-normal distribution as a limiting case.

Some work has been done on exploiting the aforementioned distributions to extend the MFA model. In fact, the work of \cite{murray14b,murray14a} and \cite{smcnicholas17} is in this direction and \cite{tortora16} develop a mixture of generalized hyperbolic factor analyzers (MGHFA) model along analogous lines. \cite{lin16} develop a mixture of factor analyzers using the multivariate skew-normal distribution used in \cite{lin09}, and \cite{murray17a} develop a mixture of factor models using the formulation of the skew-$t$ distribution employed in \cite{lin10}. The logical conclusion to this vein of research is a mixture of HTH factor analyzers (MHTHFA) model, which is developed herein. 

%Herein a mixture of hidden truncation hyperbolic factor analyzers (HTHFA) model is developed and background material related to the development of the HTHFA model is discussed in Section \ref{sec:background}. In Section \ref{sec:HTHFA} we develop the HTHFA mixture model and describe an algorithm for model fitting and parameter estimation. In Section \ref{sec:apps}, two data sets are used to illustrate clustering performance. We conclude with a discussion of the HTHFA model and future work in Section \ref{sec:discussion}. 

\section{Background}\label{sec:background}

\subsection{Mixture Model-Based Classification}

Suppose we observe $p$-dimensional $\vecx_1,\ldots,\vecx_n$ from a $G$-component finite mixture model. First, suppose all $n$ are unlabelled, i.e., a clustering scenario. Then, the model-based clustering likelihood can be written
$$\mathcal{L}(\vecvarthet)=\prod_{i=1}^n\sum_{g=1}^G\pi_gf(\vecx_i\mid\vectheta_g),$$
where $\pi_g>0$, such that $\sum_{g=1}^G\pi_g=1$, is the $g$th mixing proportion, $f(\vecx\mid\vectheta_g)$ is the $g$th component density, and $\vecvarthet=(\pi_1,\ldots,\pi_G,\vectheta_1,\ldots,\vectheta_G)$.
Now, suppose $k$ of the $n$ are unlabelled and we want to use all $n$ to find labels for the $k$ unlabelled observations, i.e., semi-supervised classification. Following \cite{mcnicholas10c}, suppose it is the first $k$ observations that are unlabelled. Then, the likelihood can be written
$$\mathcal{L}(\vecvarthet)=\prod_{i=1}^k\prod_{g=1}^G[\pi_gf(\vecx_i\mid\vectheta_g)]^{z_{ig}}\prod_{j=k+1}^n\sum_{h=1}^H\pi_gf(\vecx_j\mid\vectheta_h),$$
for $H\geq G$ ($H=G$ is commonly assumed), where $z_{ig}=1$ if $\vecx_i$ belongs to component~$g$ and $z_{ig}=0$ otherwise. This semi-supervised classification paradigm is also referred to as partial classification \citep[see][]{mclachlan92}.
Further details on model-based clustering and classification are given in the monographs by \cite{mclachlan00b} and \cite{mcnicholas16a}.

\subsection{Hidden Truncation Hyperbolic Distribution}

We generate a $p$-dimensional random variable $\vecX$ following the HTH distribution through the stochastic representation 
 $$\vecX=\vecmu+\sqrt{W}\vecY,$$
 where $\vecY\sim\text{SN}_p(\mathbf{0},\matsig,\vecLambda)$ and $W\sim\text{GIG}(\omega,\omega,\lambda)$. Here $\vecY\sim\text{SN}_p(\mathbf{0},\matsig,\vecLambda)$ denotes that $\vecY$ follows the skew-normal distribution of \cite{sahu03} with density 
 \begin{equation*}
f_{\text{\tiny SN}}(\vecy\mid\vecmu,\matsig,\vecLambda)=2^q\phi_p(\vecy\mid\vecmu,\vecOmega)\Phi_r(\vecLambda'\vecOmega^{-1}(\vecy-\vecmu)\mid\vecDelta),
\label{eq:sndensity}
\end{equation*}
with location vector $\vecmu$, scale matrix $\matsig$, $p\times r$ skewness matrix $\vecLambda$, $\vecOmega=\matsig+\vecLambda\vecLambda'$, $\vecDelta=\mathbf{I}_q-\vecLambda'\vecOmega^{-1}\vecLambda$ and where $\phi_p(\cdot\mid\vecmu,\matsig)$ 
and $\vecPhi_r(\cdot\mid\vecmu,\matsig)$ are the density and cumulative distribution function, respectively, of the multivariate normal distribution. Furthermore, $W\sim\text{GIG}(\psi,\chi,\lambda)$ denotes that $W$ follows the generalized inverse Gaussian distribution with density
$$p(w\mid\psi,\chi,\lambda)=\frac{(\psi/\chi)^{\lambda/2}w^{\lambda-1}}{2K_\lambda(\sqrt{\psi\chi})}\exp{\bigg\{-\frac{\psi w + \chi/w}{2}\bigg\}},$$
 for $w>0$ with $(\psi, \chi) \in \mathbb{R}_{+}^2$ and $\lambda\in\mathbb{R}$.
  The HTH density is
 \begin{equation*}
 \begin{split}
&f_{\text{\tiny HTH}}(\vecx\mid\vectheta)=2^qh_p(\vecx\mid\vecmu,\vecOmega,\lambda,\omega,\omega)\\
&\qquad\times H_q\left(\vecLambda'\vecOmega^{-1}(\vecx-\vecmu)\left(\frac{\omega}{\omega+\vecdelta(\vecx\mid\vecmu,\vecOmega)}\right)^{1/4}~\bigg|~\mathbf{0},\mathbf\vecDelta,\lambda-(p/2),\gamma,\gamma\right),
\label{HTHdens}
\end{split}
 \end{equation*}
 where $\vectheta=(\vecmu,\matsig,\load,\lambda,\omega)$, $\gamma=\omega \sqrt{ 1+\delta(\vecx,\vecmu\mid\vecOmega)/\omega}$, 
 \begin{equation*}
\begin{split}
h_p(\cdot\mid\vecmu,\matsig,\lambda,\omega,\omega)&= \bigg[\frac{\chi+\delta(\vecx,\vecmu\mid\matsig)}{\psi}\bigg]^{(\lambda-p/2)/2}\\&\qquad\times\frac{[\psi/\chi]^{\lambda/2}K_{\lambda-p/2}\bigg(\sqrt{\psi[\chi+\delta(\vecx,\vecmu\mid\matsig)]}\bigg)}{(2\pi)^{p/2}\mid\mathbf\Sigma\mid^{1/2}K_{\lambda}(\sqrt{\chi\psi})},
\label{eq:ghddensity}
\end{split}
\end{equation*}
where $\delta(\vecx, \vecmu\mid \matsig)=(\vecx-\vecmu)'\matsig^{-1}(\vecx-\vecmu)$ is the squared Mahalanobis distance between $\vecx$ and $\vecmu$, $h_p(\cdot\mid\vecmu,\vecOmega,\lambda,\omega,\omega)$ is the density of a $p$-dimensional symmetric hyperbolic random variable and $H_q(\cdot\mid\vecmu,\matsig,\lambda,\omega,\omega)$ is the corresponding $q$-dimensional CDF.  Note that $\vecLambda$ is a $p\times r$ skewness matrix, where $1\leq r\leq p$. Refer to \cite{murray17b} for extensive details on the HTH distribution. 

\section{HTH Factor Analysis Model}\label{sec:HTHFA}

Consider $n$ independent $p$-dimensional random variables $\vecX_1,\ldots,\vecX_n$. The HTH factor analysis model is written 
$$\vecX_i=\vecmu+\vecB \vecU_i+\vecepsilon_i,$$ for $i=1,\ldots,n$,
where $\vecmu$ is a $p$-dimensional location parameter, $\vecB$ is a $p\times q$ matrix of factor loadings, 
$\vecU_i$ is a $q$-dimensional vector of latent factors, and $\vecepsilon_i$ is a $p$-dimensional error vector. 
Note that the $\vecX_i$ are independently distributed, as are the $\vecU_i$, the $\vecX_i$ and $\vecU_i$ are independent, and $q<p$. We also have that
$$\left[ \begin{array}{c} \vecU_i\\ \vecepsilon_i \end{array} \right] \sim \text{HTH}_{q+p} \left( \left[ \begin{array}{c} -\vecA^{-1/2}\vecLambda \veca_{\lambda} \\ \mathbf{0}_p \end{array} \right],\begin{bmatrix} \vecA^{-1} & \mathbf{0}'_{q\times p} \\ \mathbf{0}_{p\times q} & \vecD \end{bmatrix}, \left[ \begin{array}{c} \vecA^{-1/2}\vecLambda  \\ \mathbf{0}_p \end{array} \right],\omega,\omega,\lambda \right),$$
where $\vecLambda$ is a $q\times r$ skewness matrix, 
$$\vecA=\mathbf{I}_q+\left[1-\veca_{\lambda}'\veca_{\lambda}\frac{K_{\lambda+1}(\omega)}{K_{\lambda}(\omega)}\right]\vecLambda\vecLambda',\qquad
\veca_{\lambda}=\sqrt{\frac{2}{\pi}}\frac{K_{\lambda+1/2}(\omega)}{K_{\lambda}(\omega)}\mathbf{1}_r,$$
$\mathbf{1}_r$ denotes an $r$-dimensional vector of 1s, and $\vecD$ is a diagonal matrix with positive diagonal elements. Note this model requires that $1\leq r\leq q$ and the value of $q$ must satisfy \begin{equation}\label{eqn:lawley}(p-q)^2>p+q.\end{equation} See \cite{lawley62} for details about \eqref{eqn:lawley}.
%We can show that $$\mathbb{E}(\vecU)=0$$ and $\text{cov}(U)=\frac{K_{\lambda+1}(\omega)}{K_{\lambda}(\omega)}I_q.$
We can show that 
$\vecX_i\mid w_i\sim \text{SN}_p(\vecmu-\vecalpha\veca_{\lambda},w_i\matsig,\sqrt{w_i}\vecalpha)$ and 
$\vecX_i\sim \text{HTH}_p(\vecmu-\vecalpha\veca_{\lambda},\matsig,\vecalpha,\lambda,\omega,\omega),$
where 
$W\sim \text{GIG}(\omega,\omega,\lambda),$
$\matsig=\vecB\vecA^{-1}\vecB'+\vecD$
and 
$\vecalpha=\vecB\vecA^{-1}\vecLambda.$

It follows that the density of the HTH factor analysis model is 
 \begin{equation}
 \begin{split}
&f_{\text{\tiny HTHFA}}(\vecx\mid\vectheta)=2^rh_p(\vecx\mid\vecmu-\vecalpha\veca_{\lambda},\vecOmega,\lambda,\omega,\omega)\times\\
&H_r\left(\vecalpha'\vecOmega^{-1}(\vecx-\vecmu+\vecalpha\veca_{\lambda})\left[\frac{\omega}{\omega+\vecdelta(\vecx\mid\vecmu-\vecalpha\veca_{\lambda},\vecOmega)}\right]^{1/4}~\bigg|~\mathbf{0},\mathbf\vecDelta,\lambda-\frac{p}{2},\gamma,\gamma\right),
\label{HTHFAdens}
\end{split}
 \end{equation}
 where $\vectheta=(\vecmu,\vecB,\vecD,\vecLambda,\omega,\lambda)$, $\vecOmega=\matsig+\vecalpha\vecalpha'$, and $\vecDelta=\textbf{I}_r-\vecalpha'\vecOmega^{-1}\vecalpha$. For notational convenience, hereafter let $\tilde{\vecB}=\vecB\vecA^{-1/2}$, $\tilde{\vecU}=\vecA^{1/2}\vecU$, and $\vecr=\vecmu-\vecalpha\veca_{\lambda}$. 
 The following hierarchical representation exists for the HTHFA model:
 \begin{equation*}
\begin{split}
\vecX\mid(\tilde{\vecu},\vecv,w)&\sim\mathcal{N}_p(\vecmu+\tilde{\vecB}\tilde{\vecU},w\vecD),\\
\tilde{\vecU}\mid\vecv,w&\sim\mathcal{N}_q\left(\vecLambda(\vecv-\veca_{\lambda}),w\mathbf{I}_q\right),\\
\vecV\mid w&\sim\text{TN}_r(\mathbf{0},w\mathbf{I}_r),\\
W&\sim\text{GIG}(\omega,\omega,\lambda).
\end{split}
\end{equation*}

\section{Mixtures of HTH Factor Analyzers}
\subsection{The Model}
We develop a MHTHFA to model high-dimensional heterogenous data. Consider $n$ independent $p$-dimensional random variables $\vecX_1,\ldots,\vecX_n$. The MTHTFA model is given by 
$$\vecX_i=\vecmu_g+\vecB_g \vecU_{ig}+\vecepsilon_{ig}$$ with probability $\pi_g$, for $i=1,\ldots,n$ and $g=1,\ldots,G$,
where $\pi_g>0$, $\sum_{g=1}^{G}\pi_g=1$, $\vecmu_g$ is a $p$-dimensional location parameter, $\vecB_g$ is a $p\times q$ matrix of factor loadings, $\vecU_{ig}$ is a $q$-dimensional vector of latent factors, and $\vecepsilon_{ig}$ is a $p$-dimensional error vector. Analogous independence relations hold for and between the $\vecX_i$ and $\vecU_i$ as for the HTH factor analysis model (Section~\ref{sec:HTHFA}). We also have that
$$\left[ \begin{array}{c} \vecU_{ig}\\ \vecepsilon_{ig} \end{array} \right] \sim \text{HTH}_{q+p} \left(\left[\begin{array}{c} -\vecA_g^{-1/2}\vecLambda_g \veca_{\lambda_g} \\ \mathbf{0}_p \end{array} \right],\begin{bmatrix} \vecA_g^{-1} & \mathbf{0}'_{q\times p} \\ \mathbf{0}_{p\times q} & \vecD_g \end{bmatrix}, \left[\begin{array}{c} \vecA_g^{-1/2}\vecLambda_g  \\ \mathbf{0}_p \end{array} \right],\omega_g,\omega_g,\lambda_g \right),$$
where $\vecLambda_g$ is a $q\times r$ skewness matrix, 
$$\vecA_g=\mathbf{I}_q+\left[1-\veca_{\lambda_g}'\veca_{\lambda_g}\frac{K_{\lambda_g+1}(\omega_g)}{K_{\lambda_g}(\omega_g)}\right]\vecLambda_g\vecLambda_g',\qquad
\veca_{\lambda_g}=\sqrt{\frac{2}{\pi}}\frac{K_{\lambda_g+1/2}(\omega_g)}{K_{\lambda_g}(\omega_g)}\mathbf{1}_r,$$
$\mathbf{1}_r$ denotes an $r$-dimensional vector of 1s, and $\vecD_g$ is a diagonal matrix with positive diagonal elements. Again, we require that $1\leq r\leq q$ and the value of $q$ must satisfy \eqref{eqn:lawley}.
The density of the MHTHFA model is 
\begin{equation}
g(\vecx\mid\vecvarthet)=\sum_{g=1}^{G}\pi_g f_\text{\tiny HTHFA}(\vecx\mid\vectheta_g),
\label{eq:generalmixture1}
\end{equation}
where 
$\vecvarthet=(\pi_1,...,\pi_G,\vectheta_1,...,\vectheta_G)$, 
$\pi_g$ is the $g$th mixing proportion, $\vectheta_g=(\vecmu_g,\matsig_g,\vecLambda_g,\lambda_g,\omega_g,\omega_g)$ and $f_\text{\tiny HTHFA}(\vecx\mid\vectheta_g)$ is as defined in \eqref{HTHFAdens}. 

\subsection{ECM algorithm for MHTHFA}

We employ an expectation conditional maximization (ECM) algorithm \citep{meng93} for model fitting and parameter estimation. The complete-data log-likelihood for the HTHFA model is 
\begin{equation*}
\begin{split}
l_c&=C+\sum^{G}_{g=1}\Bigg[(\lambda_g-1)\sum^{n}_{i=1}\log w_{ig}-n\log K_{\lambda_g}(\omega_g)-\frac{\omega_g}{2}\sum^{n}_{i=1}\left(w_{ig}+\frac{1}{w_{ig}}\right)\\
&-\frac{1}{2}\sum^{n}_{i=1}\frac{1}{w_{ig}}\Big[ \tilde{\vecu}_{ig}'\tilde{\vecu}_{ig}-\tilde{\vecu}_{ig}'\vecLambda_g(\vecv_{ig}-\veca_{\lambda_g})-(\vecv_{ig}-\veca_{\lambda_g})'\vecLambda_g'\tilde{\vecu}_{ig}\\&+(\vecv_{ig}-\veca_{\lambda_g})'\vecLambda_g'\vecLambda_g(\vecv_{ig}-\veca_{\lambda_g})\Big]
-\frac{n}{2}\log|\vecD_g|-\frac{1}{2}\tr\left\{\vecD_g^{-1}\sum^{N}_{i=1}\tilde{\noisev}_{ig}\right\}\Bigg]
\end{split}
\end{equation*}
where 
$$\tilde{\noisev}_{ig}=\frac{1}{w_{ig}}(\vecx_i-\vecmu_g-\tilde{\vecB}_g\tilde{\vecu}_{ig})(\vecx_i-\vecmu_g-\tilde{\vecB}_g\tilde{\vecu}_{ig})',$$ $C$ is a constant with respect to the model parameters, and $z_{ig}$ is as defined before. %From this complete-data log-likelihood function, that the maximum likelihood estimates of all model parameters are derived. 
The algorithm alternates between a expectation (E) step and a conditional-maximization (CM) step. On the E-step, we compute the expected value of the complete-data log-likelihood conditional on the current parameter estimates. For the MHTHFA model, this involves the following conditional expectations:
\begin{equation*}
\begin{split}
&\mathbb{E}\left[W_{ig}\mid\vecx_i, z_{ig}=1\right]\equalscolon\hat{a}_{ig},\quad
\mathbb{E}\left[{1}/{W_{ig}}\mid\vecx_i,z_{ig}=1\right]\equalscolon\hat{b}_{ig}\\
&\mathbb{E}[\log W_{ig}\mid\vecx_i,z_{ig}=1]\equalscolon\hat{c}_{ig},\quad
\mathbb{E}[(1/W_{ig})\tilde{\vecU}_{ig}\mid\vecx_i,z_{ig}=1]\equalscolon\hat{\vecs}_{1ig}\\
&\mathbb{E}[(1/W_{ig})\tilde{\vecU}_{ig}\tilde{\vecU}_{ig}'\mid\vecx_i,z_{ig}=1]\equalscolon\hat{\vecs}_{2ig},\quad
\mathbb{E}[(1/W_{ig})\vecV_{ig}\mid\vecx_i,z_{ig}=1]\equalscolon\hat{\vecs}_{3ig}\\
&\mathbb{E}[(1/W_{ig})\vecV_{ig}\vecV_{ig}'\mid\vecx_i,z_{ig}=1]\equalscolon\hat{\vecs}_{4ig},\ \
\mathbb{E}[(1/W_{ig})\vecV_{ig}\tilde{\vecU}_{ig}'\mid\vecx_i,z_{ig}=1]\equalscolon\hat{\vecs}_{5ig}\\
\end{split}
\end{equation*}
Details on these expectations are given in the appendix. 

At each CM-step, the following model parameters are updated sequentially and conditionally on the other parameters. The mixing proportions and location parameters are updated via
\begin{equation}
\hat{\pi}_g=\frac{n_g}{n}\qquad\text{and}\qquad
%\end{equation}
%and the $g^{th}$ location parameter by
%\begin{equation}
\hat{\vecmu}_g=\frac{\sum^{n}_{i=1}\hat{z}_{ig}(\hat{b}_{ig}\vecx_i-\hat{\tilde{\vecB}}_g\hat{\vecs}_{1ig})}{\sum^{n}_{i=1}\hat{z}_{ig}\hat{b}_{ig}},
\end{equation}
respectively, where $n_g=\sum^{n}_{i=1}\hat{z}_{ig}$. We update $\tilde{\vecB}_g$ by
\begin{equation}
\hat{\tilde{\vecB}}_g=\left[\sum^{n}_{i=1}\hat{z}_{ig}\left( \vecx_i-\hat{\vecmu}_g\right)\hat{\vecs}_{1ig}'\right]\left[ \sum^{n}_{i=1}\hat{z}_{ig}\hat{\vecs}_{2ig}\right]^{-1}
\end{equation}
and the update for $\vecD_g$ is
\begin{equation}
 \hat{\vecD}_g=\frac{1}{n_g}\text{diag}\left\{\sum^{n}_{i=1}\hat{z}_{ig}\hat{\noisev}_{ig}\right\},
 \end{equation}
 and
 $$\hat{\vecPsi}_{ig}=\hat{b}_{ig}(\vecx_i-\hat{\vecmu}_g)(\vecx_i-\hat{\vecmu}_g)'-(\vecx_i-\hat{\vecmu}_g)(\tilde{\vecB}_g\hat{\vecs}_{1ig})'-(\tilde{\vecB}_g\hat{\vecs}_{1ig})(\vecx_i-\hat{\vecmu}_g)'+\hat{\tilde{\vecB}}_g\hat{\vecs}_{2ig}\hat{\tilde{\vecB}}_g'.$$
The update for the $\vecLambda_g$ is
\begin{equation}
\hat{\vecLambda}_g=\sum^{n}_{i=1}\left(\hat{\vecs}_{5ig}-\hat{\vecs}_{1ig}\veca'_{\lambda_g}\right)\left(\hat{\vecs}_{4ig}-\veca_{\lambda_g}\hat{s}'_{3ig}-\hat{\vecs}_{3ig}\veca_{\lambda_g}'-\hat{b}_{ig}\veca_{\lambda_g}'\veca_{\lambda_g}\right)^{-1},
\end{equation}
and the update for $\omega_g$ is given by
$$\hat{\omega}_g= \omega_g -  \frac{  \partial_{\omega}  t }{ \partial^2_{\omega}  t }\bigg|_{\omega=\omega_g},$$
where
$$t_g(\omega_g,\lambda_g) = -\log K_{\lambda_g} \left(\omega_g \right) + (\lambda_g-1)\overline{c}_g - \frac{\omega_g}{2} \left( \overline{a}_g + \overline{b}_g \right),$$
$\overline{a}_g=\sum^{n}_{i=1}\hat{z}_{ig}\hat{a}_{ig}/n_g$, $\overline{b}_g=\sum^{n}_{i=1}\hat{z}_{ig}\hat{b}_{ig}/n_g$, and $\overline{c}_g=\sum^{n}_{i=1}\hat{z}_{ig}\hat{c}_{ig}/n_g$.
We update $\lambda_g$ by
$\hat{\lambda}_g=\bar{c}_g\lambda_g/m_g$,
where
\begin{equation*}
\begin{split}
m_g=&\frac{\partial}{\partial \lambda}\log K_\lambda(\hat{\omega}_g)\bigg|_{\lambda=\lambda_g}+\frac{1}{2}\frac{\partial}{\partial \lambda}\frac{K_{\lambda+1/2}(\hat{\omega}_g)}{K_\lambda(\hat{\omega}_g)}\bigg|_{\lambda=\lambda_g}\Bigg[\sum^{n}_{i=1}\text{tr}\left\{\hat{\vecs}_{1ig}\sqrt{\frac{2}{\pi}}\mathbf{1}_r\right\}\\
&+ \sum^{n}_{i=1}\text{tr}\left\{\sqrt{\frac{2}{\pi}}\mathbf{1}_r\hat{\vecs}_{1ig}'\hat{\vecLambda}_g\right\}- \sum^{n}_{i=1}\text{tr}\left\{\Bigg(\hat{\vecs}_{3ig}\sqrt{\frac{2}{\pi}}\mathbf{1}'_r+\sqrt{\frac{2}{\pi}}\mathbf{1}_r\hat{\vecs}_{3ig}\Bigg)\hat{\vecLambda}_g'\hat{\vecLambda}_g\right\}\Bigg]\\
&-\frac{1}{2}\frac{\partial}{\partial \lambda}\left(\frac{K_{\lambda+1/2}(\hat{\omega}_g)}{K_\lambda(\hat{\omega}_g)}\right)^2\bigg|_{\lambda=\lambda_g}\sum^{n}_{i=1}\text{tr}\left\{\hat{b}_{ig}\sqrt{\frac{2}{\pi}}\mathbf{1}_r\sqrt{\frac{2}{\pi}}\mathbf{1}'_r\hat{\vecLambda}_g'\hat{\vecLambda}_g \right\}.
\end{split}
\end{equation*}

\subsection{Model Selection}
The values of $q$ and $r$ will need to be selected as well as, perhaps, the number of components $G$. The Bayesian information criterion \citep[BIC;][]{schwarz78} is the most popular model selection criterion for model-based clustering and semi-supervised classification. The BIC can be written
$$
\text{BIC}=2l_{\text{\tiny obs}}(\hat\vecvarthet)-\rho\log n,
$$
where $l_{\text{\tiny obs}}(\hat\vecvarthet)$ is the maximized observed likelihood and $\rho$ is the number of free parameters in the model. Note that, for our MHTHFA model,
$$\rho=G-1+G\left[p+qr+2+pq+p-\frac{q}{2}(q-1)\right].$$

\subsection{Initialization and Convergence}

In our ECM algorithms, the group memberships $\hat{z}_{ig}$ are initialized using $k$-means clustering. For each ECM algorithm, five sets of starting values are obtained by performing $k$-means clustering and our ECM algorithm is initialized using %as described herein. For the five sets of starting values, the log-likelihood was calculated and 
the set of values that corresponds the largest log-likelihood value for the MHTHFA model in question.

The parameters $\hat{\vecmu}_g$ and $\hat{\matsig}_g$ are initialized using a weighted mean and covariance matrix, respectively. The matrix $\hat{\vecLambda}_g$ is initialized using values randomly generated from a normal distribution with mean 0 and standard deviation 1, and $\hat{\omega}_g$ and $\hat{\lambda}_g$ are each initialized as 1. We initialize the matrix of factor loadings $\hat{\vecB}_g$ following the approach outlined by \cite{mcnicholas08}.

We determine convergence of the ECM algorithm using a criterion based on the Aitken acceleration \citep{aitken26}. The Aitken acceleration at iteration $k$ is $$a^{(k)}=\frac{l^{(k+1)}-l^{(k)}}{l^{(k)}-l^{(k-1)}},$$ where $l^{(k)}$ is the log-likelihood at iteration $k$. An asymptotic estimate of the log-likelihood at iteration $k$ is $$l_{\infty}^{(k)}=l^{(k-1)}+\frac{1}{1-a^{(k-1)}}(l^{(k)}-l^{(k-1)}).$$ Following \cite{lindsay95}, we stop the algorithm when $l_{\infty}^{(k)}-l^{(k)}<\epsilon$, with $\epsilon=0.01$, for the analyses herein (Section~\ref{sec:apps}).

\section{Illustrations}\label{sec:apps}

\subsection{Overview}

The MHTHFA model is illustrated for clustering (Section~\ref{sec:ais}) and semi-supervised classification (Section~\ref{sec:sonar}) using two well-known datasets. Because the true classes of the points that are treated as unlabelled are actually known, performance can be assessed using the adjusted Rand index \citep[ARI;][]{hubert85}. The ARI takes a value 1 for perfect class agreement and has expected value 0 under random classification. Negative values of the ARI are also possible and reflect classification performance that is, in some sense, worse than guessing. Extensive details on the ARI are given by \cite{steinley04}.

\subsection{Australian Institute of Sport Data}\label{sec:ais}

We consider the Australian Institute of Sport (AIS) data which contains 11 continuous variables for 100 female and 102 male athletes. These data are available in the UCI Machine Learning Repository \citep{lichman13}. The MHTHFA model is fitted for $G=2$, $q=1,\ldots,6$, $r=1,2,3$, and $r\leq q$. For comparison, we also fit the MGHFA and PGMM models for $G=2$ and $q=1,\ldots,6$. The BIC is used to select the best model in each case and the results are summarized in Table \ref{tab:HTHFA_ais}. Although all models obtain very good clustering results, the selected MHTHFA model outperforms the MGHFA and PGMM approaches. 
\begin{table}[ht]
\caption{Results from the best MHTHFA, MGHFA, and PGMM models fit to the AIS data for $G=2$, as selected by the BIC.}
\centering
\begin{tabular*}{0.6\textwidth}{@{\extracolsep{\fill}}lrrrr}
\hline
Model & $q$ & $r$ & BIC & ARI\\ 
\hline
MHTHFA&4&2&$ -2369.3$&0.92 \\
MGHFA &4&&$-2266.6$&0.90\\
PGMM &4&&$ -2394.8$&0.81\\
\hline
\end{tabular*}
\label{tab:HTHFA_ais}
\end{table}

\subsection{Sonar Data}\label{sec:sonar}

\cite{gorman88} report data on patterns obtained by bouncing sonar signals off a metal cylinder and rocks, respectively, at various angles and under various conditions. These data are available from the UCI machine learning repository. In all, 111 signals are obtained by bouncing sonar signals off a metal cylinder and 97 are obtained by bouncing sonar signals off rocks. To illustrate the MHTHFA approach for semi-supervised classification, 104 of the 208 patterns are randomly selected to be treated as unlabelled. This results in 53 of the metal signals and 51 of the rock signals being treated as unlabelled. The MHTHFA, MGHFA, and PGMM approaches are fitted to these data, for semi-supervised classification, for $G=2$, $q=1,\ldots,8$ and, for MHTHFA, $r=1,2$. The classification results, based on the best model selected by the BIC in each case, are given in Table~\ref{tab:sonar}. The MHTHFA model substantially outperforms both the MGHFA and PGMM approaches (Tables~\ref{tab:sonar} and~\ref{tab:HTHFA_sonar}).
\begin{table}[ht]
\caption{Cross-tabulations of true (metal, rock) against predicted (A, B) classes for the selected MHTHFA, MGHFA, and PGMM models on the sonar data.}
\centering
\begin{tabular*}{0.7\textwidth}{@{\extracolsep{\fill}}lrrrrrrrr}
\hline
&\multicolumn{2}{c}{MHTHFA}&&\multicolumn{2}{c}{MGHFA}&&\multicolumn{2}{c}{PGMM}\\ 
\cline{2-3}\cline{5-6}\cline{8-9}
&A&B&&A&B&&A&B\\ 
\hline
Metal &48&5&&49&4&&42&11\\
Rock &7&44&&31&20&&24&27\\
\hline
\end{tabular*}
\label{tab:sonar}
\end{table}
\begin{table}[ht]
\caption{Results from the best MHTHFA, MGHFA, and PGMM models fit to the sonar data for $G=2$, as selected by the BIC.}
\centering
\begin{tabular*}{0.6\textwidth}{@{\extracolsep{\fill}}lrrrr}
\hline
Model & $q$ & $r$ & BIC & ARI\\ 
\hline
MHTHFA&7&1&$-27258.8$&0.59 \\
MGHFA &6&&$-27627.6$&0.05\\
PGMM &7&&$-29279.9$&0.10\\
\hline
\end{tabular*}
\label{tab:HTHFA_sonar}
\end{table}

%Sonar with 50% unlabelled. i.ei. 104 out of 208 unlabelled with G=2
%
%PGMM q=7 
%     1  2
%  1 42 24
%  2 11 27
%
%MHTHFA, q=8   
%     1  2
%  1 50  9
%  2  3 42
%
%MGHFA q=2
%1  2
%1 49 31
%2  4 20
%

\section{Discussion}\label{sec:discussion}

The MFA model has been extended using the HTH distribution that was developed by \cite{murray17b}. The HTH distribution contains many of the non-Gaussian distributions used in the model-based clustering literature as special and limiting cases. The resulting MHTHFA model retains the flexibility of the HTH mixture model with the added advantage of being able to model high-dimensional data. This work can be viewed as completing a line of research on extensions to the MFA model using non-Gaussian distributions based on hidden truncation. 
Illustrations on well known datasets demonstrate that the MHTHFA model is effective for clustering and semi-supervised classification; in fact, it outperforms both the MGHFA and PGMM approaches. Given the flexibility of this model and the inherent ability to capture various distributions as special cases, this model is particularly useful for high dimensional clustering applications where the underlying distribution is unknown. 

Future work will focus on decreasing the computation time required to fit the MHTHFA model.  The $\sf{R}$ programming language \citep{R15} has been used for all model implementation to date but developing parallel code using python or julia will reduce the overall computation time and increase the practical applicibility of this model. 
Other work could focus on developing a parsimonious family of MHTHFA models analogous to the PGMM models; however, more efficient implementation is essentially a pre-requisite for such developments. An analogous approach to that presented herein could be taken to extending the mixture of common factor analyzers model \citep{yoshida04,baek10} to the mixture of HTH distributions; the resulting approach might be effective for clustering higher dimensional data. The same may be true of an MHTHFA analogue of the LASSO-penalized approach of \cite{bhattacharya14a}. Consideration of how the MTHTFA model works within the fractionally supervised paradigm may also be of interest \citep[see][]{vrbik15,gallaugher18b}. Finally, matrix variate analogues of the HTH distribution and MTHTFA model will be considered and may follow somewhat similar lines to the non-Gaussian matrix variate mixture work of \cite{gallaugher17,gallaugher18a,gallaugher18c}.

%\newpage

\appendix
\section{E-step Calculations}

Herein we present the expectations required for the E-step of the ECM algorithm for the mixtures of HTH factor analyzers model.

\subsection{$\mathbb{E}[W_{ig}\mid \vecx_i,z_{ig}=1]$ and $\mathbb{E}[1/W_{ig}\mid \vecx_i,z_{ig}=1]$ }

To derive the expectation $\mathbb{E}[W_{ig}\mid \vecx_i,z_{ig}=1]$ and $\mathbb{E}[1/W_{ig}\mid \vecx_i,z_{ig}=1]$ as well as $\mathbb{E}[\log W_{ig}\mid \vecx_i,z_{ig}=1]$ in the following section, first note that 
\begin{equation}
\begin{split}
f(&w_{ig}\mid\vecx_i,z_{ig}=1)
=\frac{w^{\lambda_g-p/2-1}}{2K_{\lambda_g-p/2}(\sqrt{\omega_g(\omega_g+\delta(\vecx_i\mid\vecr_g,\vecOmega_g))})}\left[ \frac{\omega_g}{\omega_g+\delta(\vecx_i\mid\vecr_g,\vecOmega_g)}\right]^{(\lambda_g-p/2)/2}\\
&\times\exp\left\{\omega_g w+\frac{\omega_g+\delta_g(\vecx_i\mid\vecr_g,\vecOmega_g)}{w}\right\}\Phi\left(\vecalpha_g'\vecOmega_g^{-1}(\vecx_i-\vecr_g)/\sqrt{w}\mid\vecDelta_g\right)\\
&\div H_r\left(\vecalpha_g'\vecOmega_g^{-1}(\vecx_i-\vecr_g)\left(\frac{\omega_g}{\omega_g+\vecdelta(\vecx_i\mid\vecr_g,\vecOmega_g)}\right)^{1/4}\bigg|\mathbf{0},\vecDelta_g,\lambda_g-(p/2),\gamma_g,\gamma_g\right).
\end{split}         
\end{equation}
Therefore, 
\begin{equation*}
\begin{split}
\mathbb{E}&\left[W_{ig}\mid\vecx_i,z_{ig}=1 \right]\\
&=\int^{\infty}_0\frac{w^{\lambda_g-p/2}}{2K_{\lambda_g-p/2}(\sqrt{\omega_g(\omega_g+\delta(\vecx_i\mid\vecr_g,\vecOmega_g))})}\left[ \frac{\omega_g}{\omega_g+\delta(\vecx_i\mid\vecr_g,\vecOmega_g)}\right]^{(\lambda_g-p/2)/2}\\
&\qquad\times\exp\left\{\omega_g w+\frac{\omega_g+\delta(\vecx_i\mid\vecr_g,\vecOmega_g)}{w}\right\}\Phi\left(\vecalpha_g'\vecOmega_g^{-1}(\vecx_i-\vecr_g)/\sqrt{w}\mid\vecDelta_g\right)\\
&\qquad\div H_r\left(\vecalpha_g'\vecOmega_g^{-1}(\vecx_i-\vecr_g)\left(\frac{\omega_g}{\omega_g+\vecdelta(\vecx_i\mid\vecr_g,\vecOmega_g)}\right)^{1/4}\bigg|\mathbf{0},\mathbf\vecDelta_g,\lambda_g-(p/2),\gamma_g,\gamma_g\right)dw\\
&=\frac{K_{\lambda_g-p/2+1}(\sqrt{\omega_g(\omega_g+\delta(\vecx_i\mid\vecr_g,\vecOmega_g))})}{K_{\lambda_g-p/2}(\sqrt{\omega_g(\omega_g+\delta(\vecx_i\mid\vecr_g,\vecOmega_g))})}
\left[ \frac{\omega_g+\delta(\vecx_i\mid\vecr_g,\vecOmega_g)}{\omega_g}\right]^{1/2}\\
&\qquad\times H_r\left(\vecalpha_g'\vecOmega_g^{-1}(\vecx_i-\vecr_g)\left(\frac{\omega_g}{\omega_g+\vecdelta(\vecx_i\mid\vecr_g,\vecOmega_g)}\right)^{1/4}\bigg|\mathbf{0},\mathbf\vecDelta_g,\lambda_g-(p/2)+1,\gamma_g,\gamma_g\right)\\
&\qquad\div H_q\left(\vecalpha_g'\vecOmega_g^{-1}(\vecx_i-\vecr_g)\left(\frac{\omega_g}{\omega_g+\vecdelta(\vecx_i\mid\vecr_g,\vecOmega_g)}\right)^{1/4}\bigg|\mathbf{0},\mathbf\vecDelta_g,\lambda_g-(p/2),\gamma_g,\gamma_g\right),
\end{split}         
\end{equation*}
\begin{equation*}
\begin{split}
\mathbb{E}&\left[1/W_{ig}\mid\vecx_i,z_{ig}=1 \right]\\
&=\int^{\infty}_0\frac{w^{\lambda_g-p/2-2}}{2K_{\lambda_g-p/2}(\sqrt{\omega_g(\omega_g+\delta(\vecx_i\mid\vecr_g,\vecOmega_g))})}\left[ \frac{\omega_g}{\omega_g+\delta(\vecx_i\mid\vecr_g,\vecOmega_g)}\right]^{(\lambda_g-p/2)/2}\\
&\qquad\times\exp\left\{\omega_g w+\frac{\omega_g+\delta(\vecx_i\mid\vecr_g,\vecOmega_g)}{w}\right\}\Phi\left(\vecalpha_g'\vecOmega_g^{-1}(\vecx_i-\vecr_g)/\sqrt{w}\mid\vecDelta_g\right)\\
&\qquad\div H_r\left(\vecalpha_g'\vecOmega_g^{-1}(\vecx_i-\vecr_g)\left(\frac{\omega_g}{\omega_g+\vecdelta(\vecx_i\mid\vecr_g,\vecOmega_g)}\right)^{1/4}\bigg|\mathbf{0},\mathbf\vecDelta_g,\lambda_g-(p/2),\gamma_g,\gamma_g\right)dw\\
&=\frac{K_{\lambda_g-p/2-1}(\sqrt{\omega_g(\omega_g+\delta(\vecx_i\mid\vecr_g,\vecOmega_g))})}{K_{\lambda_g-p/2}(\sqrt{\omega_g(\omega_g+\delta(\vecx_i\mid\vecr_g,\vecOmega_g))})}
\left[ \frac{\omega_g+\delta(\vecx_i\mid\vecr_g,\vecOmega_g)}{\omega_g}\right]^{-1/2}\\
&\qquad\times H_r\left(\vecalpha_g'\vecOmega_g^{-1}(\vecx_i-\vecr_g)\left(\frac{\omega_g}{\omega_g+\vecdelta(\vecx_i\mid\vecr_g,\vecOmega_g)}\right)^{1/4}\bigg|\mathbf{0},\mathbf\vecDelta_g,\lambda_g-(p/2)-1,\gamma_g,\gamma_g\right)\\
&\qquad\div H_q\left(\vecalpha_g'\vecOmega_g^{-1}(\vecx_i-\vecr_g)\left(\frac{\omega_g}{\omega_g+\vecdelta(\vecx_i\mid\vecr_g,\vecOmega_g)}\right)^{1/4}\bigg|\mathbf{0},\mathbf\vecDelta_g,\lambda_g-(p/2),\gamma_g,\gamma_g\right).
\end{split}         
\end{equation*}

%%%%%%%%%%%%%%%%%%%%%%%%%%%%%%%%%%%%%%%%%%%%%%%%%%%
%%%%%%%%%%%%%%%%%%%%%%%%%%%%%%%%%%%%%%%%%%%%%%%%%%%

\subsection{$\mathbb{E}[\log W_{ig}\mid \vecx_i,z_{ig}=1]$}

To update $\mathbb{E}[\log W_{ig}\mid\vecx_i,z_{ig}=1]$, where $W_{ig}\sim \text{GIG}(\psi_g,\chi_g,\lambda_g)$, first note that
\begin{equation*} 
\mathbb{E}[ \log W_{ig}\mid z_{ig}=1] 
=  \frac{  \mathrm{d} }{ \mathrm{d} \lambda } \log K_{\lambda} \left(  \sqrt{ \chi_g \psi_g } \right) + \log \left( \sqrt{ \frac{\chi_g}{\psi_g} } \right).
\end{equation*}
We can show that 
%$$W_{ig}|\vecX_i,\vecU_{ig},Z_{ig}=1\sim \text{GIG}(\psi^*=\omega_g,\chi^*=\omega_g + (\vecv_{ig}-\veck_g)'\vecDelta_g^{-1}(\vecv_{ig}-\veck_g)+\vecdelta(\vecx_i|\vecr_g,\vecOmega_g),\tau=\lambda_g-(p+r)/2)),$$ 
$$W_{ig}\mid\vecx_i,\vecv_{ig},z_{ig}=1\sim \text{GIG}(\omega_g,\omega_g + (\vecv_{ig}-\veck_g)'\vecDelta_g^{-1}(\vecv_{ig}-\veck_g)+\vecdelta(\vecx_i\mid\vecr_g,\vecOmega_g),\lambda_g-(p+r)/2)),$$ 
where $\vecr_g=\vecmu_g-\vecalpha_g\veca_{\lambda_g}$ and $\veck_g=\vecLambda'_g\vecOmega_g^{-1}(\vecx_i-\vecmu_g)$. Therefore, 
\begin{equation*} 
\mathbb{E}[\log W_{ig}\mid\vecx_i,\vecv_{ig},z_{ig}=1] 
=  \frac{  \mathrm{d} }{ \mathrm{d} \tau } \log K_{\tau} \left(  \sqrt{ \chi^* \psi^* } \right) + \log \left( \sqrt{ \frac{\chi^*}{\psi^*} } \right).
\end{equation*}
Let
\begin{equation*}
\zeta_{ig} = \sqrt{ 1 + \frac{\delta\left(\vecx_i\mid \vecmu_g, \matsig_g\right)  + (\vecv_{ig} - \veck_g)' \vecDelta_g^{-1} (\vecv_{ig} - \veck_g) }{  \omega_g } }, \end{equation*}
then $\zeta_{ig}\ge1$ and 
%$W_{ig}|X=x_i, U=u_{ig},Z_{ig}=1\sim \text{GIG}( \psi^*= \omega_g, \chi^*= \omega_g z^2, \tau =\tau )$. 
$W_{ig}\mid\vecx_i, \vecv_{ig},z_{ig}=1\sim \text{GIG}(\omega_g, \omega_g \zeta_{ig}^2, \tau )$. 
Consequently,
\begin{equation*} \mathbb{E}[ \log W_{ig} \mid \vecx_i, \vecv_{ig},z_{ig}=1] =   \frac{  \mathrm{d} }{ \mathrm{d} \tau } \log K_{\tau} \left( \omega_g \zeta_{ig}\right) + \log\zeta_{ig}.
\end{equation*}
The reader is directed to the supplementary material in \cite{murray17b} for details on a method for estimating this expectation via a series expansion. 

%%%%%%%%%%%%%%%%%%%%%%%%%%%%%%%%%%%%%%%%%%%%%%%%%%%%
%%%%%%%%%%%%%%%%%%%%%%%%%%%%%%%%%%%%%%%%%%%%%%%%%%%%

\subsection{$\mathbb{E}[(1/W_{ig})\vecV_{ig}\mid \vecx_i,z_{ig}=1]$ and $\mathbb{E}[(1/W_{ig})\vecV_{ig}\vecV_{ig}'\mid \vecx_i,z_{ig}=1]$}

Recall that $\vecV_{ig}\mid w_{ig},z_{ig}=1\sim\text{HN}_r(w_{ig}\mathbf{I}_r)$. We can show that 
\begin{equation}
\begin{split}
f(\vecv_{ig}\mid\vecx_i,z_{ig}=1)=\frac{1}{c_\lambda} h_r\left(\vecv_{ig}~\Bigg|~\veck_g,\sqrt{\frac{\omega_g+\vecdelta(\vecx_i\mid\vecr_g,\vecOmega_g)}{\omega_g}}\vecDelta_g,\lambda_g-\frac{p}{2},\gamma_g,\gamma_g\right),
\label{ux}
\end{split}
\end{equation}
where the support of $\vecV_{ig}$ is $\mathbb{R}_+^r$, i.e., the positive plane of $\mathbb{R}_r$ and $$c_\lambda= H_r\left(\veck\left(\frac{\omega}{\omega+\vecdelta(\vecx_i\mid\vecr_g,\vecOmega_g)}\right)^{1/4}\bigg|\mathbf{0},\mathbf\vecDelta_g,\lambda_g-\frac{p}{2},\gamma_g,\gamma_g \right).$$
It follows that 
$$\vecV_{ig}\mid w_{ig},\vecx_i,z_{ig}=1 \sim \text{TH}_r\left(\veck_g, \sqrt{\frac{\omega_g+\vecdelta(\vecx_i\mid\vecr_g,\vecOmega_g)}{\omega_g}}\vecDelta_g,\lambda_g-\frac{p}{2}, \gamma_g,\gamma_g);\mathbb{R}_+^r\right).$$  Here, $\text{TH}_r(\vecmu,\matsig, \lambda,\psi,\chi;\mathbb{R}_+^r)$ denotes the $r$-dimensional symmetric truncated hyperbolic distribution with density 
$$f_{\text{\tiny TH}}(\vecv\mid\vecmu,\matsig,\lambda,\psi,\chi;\mathbb{R}_+^r)=
\frac{h_r(\vecv\mid\vecmu,\matsig,\lambda,\psi,\chi)}{\int^{\infty}_{0}\ldots \int^{\infty}_{0}h_r(\vecv\mid\vecmu,\matsig,\lambda,\psi,\chi)d\vecv}\mathbb{I}_{\mathbb{R}_+^v}(\vecv),$$
and $\mathbb{I}_{\mathbb{R}_+^r}(\vecu)=1$ when
$\vecv\in \mathbb{R}_+^r$ and 0 otherwise. In this way, the symmetric hyperbolic distribution is truncated to exist only within with region $\mathbb{R}_+^r$.
To update $\mathbb{E}[(1/W_{ig})\vecV_{ig}\mid \vecx_i,z_{ig}=1]$ and $\mathbb{E}[(1/W_{ig})\vecV_{ig}\vecV_{ig}'\mid \vecx_i,z_{ig}=1]$, we can make use of the fact that 
$$\mathbb{E}[(1/W_{ig})\vecV_{ig}\mid \vecx_i,z_{ig}=1]=\mathbb{E}[(1/W_{ig})\mid \vecx_i,z_{ig}=1]\mathbb{E}[\vecY_{ig}\mid \vecx_i,z_{ig}=1]$$
and 
$$\mathbb{E}[(1/W_{ig})\vecV_{ig}\vecV_{ig}'\mid \vecx_i,z_{ig}=1]=\mathbb{E}[(1/W_{ig})\mid \vecx_i,z_{ig}=1]\mathbb{E}[\vecY_{ig}\vecY_{ig}'\mid \vecx_i,z_{ig}=1],$$
where
$$\vecY_{ig}\mid w_{ig},\vecx_i ,z_{ig}=1\sim \text{TH}_r\left(\veck_g, \sqrt{\frac{\omega_g+\vecdelta(\vecx_i\mid\vecr_g,\vecOmega_g)}{\omega_g}}\vecDelta_g,\lambda_g-\frac{p}{2}-1, \gamma_g,\gamma_g;\mathbb{R}_+^r\right).$$ 
The expectations $\mathbb{E}[\vecY_{ig}\mid \vecx_i,z_{ig}=1]$ and $\mathbb{E}[\vecY_{ig}\vecY_{ig}'\mid \vecx_i,z_{ig}=1]$ can easily be estimated using the moments of the truncated symmetric hyperbolic distribution defined in \cite{murray17b}.

%%%%%%%%%%%%%%%%%%%%%%%%%%%%%%%%%%%%%%%%%%%%%%%%%%%%%%
%%%%%%%%%%%%%%%%%%%%%%%%%%%%%%%%%%%%%%%%%%%%%%%%%%%%%%

\subsection{$\mathbb{E}[(1/W_{ig})\tilde{\vecU}_{ig}\mid \vecx_i,z_{ig}=1]$ and $\mathbb{E}[(1/W_{ig})\tilde{\vecU}_{ig}\tilde{\vecU}_{ig}'\mid \vecx_i,z_{ig}=1]$}

Note that $\tilde{\vecU}_{ig}\mid\vecx_i,\vecv_{ig},w_{ig},z_{ig}=1\sim \mathcal{N}_q(\vecq,w_{ig}\vecC)$ where $\vecq=\vecC[\vecd+\vecLambda_g(\vecV_{ig}-\veca_{\lambda_g})]$, $\vecd=\tilde{\vecB}_g'\vecD_g^{-1}(\vecX_i-\vecmu_g)$, and $\vecC=(\mathbf{I}_q+\tilde{\vecB}_g'\vecD_g^{-1}\tilde{\vecB}_g)^{-1}$. We can show
\begin{equation*}
\begin{split}
\mathbb{E}&[\tilde{\vecU}_{ig}\mid\vecx_i,z_{ig}=1]
=\mathbb{E}\{\mathbb{E}[\tilde{\vecU}_{ig}\mid\vecx_i,\vecv_{ig},w_{ig},z_{ig}=1]\mid\vecx_i ,z_{ig}=1\}\\
&\qquad=\mathbb{E}\{\vecC[ \vecd+\vecLambda_g(\vecV_{ig}-\veca_{\lambda_g})]\mid\vecx_i,z_{ig}=1\}
=\vecC\{ \vecd+\vecLambda_g(\mathbb{E}[\vecV_{ig}\mid\vecx_i,z_{ig}=1]-\veca_{\lambda_g})\}.\\
\end{split}         
\end{equation*}
Therefore, it follows that
\begin{equation*}
\begin{split}
\mathbb{E}&[(1/W_{ig})\tilde{\vecU}_{ig}\mid\vecx_i,z_{ig}=1]
=\mathbb{E}\{\mathbb{E}[(1/W_{ig}) \tilde{\vecU}_{ig}\mid\vecx_i,\vecv_{ig},w_{ig},z_{ig}=1]\mid\vecx_i,z_{ig}=1 \}\\
&=\mathbb{E}\{(1/W_{ig})[ \vecC\vecd+\vecC\vecLambda_g(\vecV_{ig}-\veca_{\lambda_g})]\mid\vecx_i,z_{ig}=1\}\\
&=\vecC\{\vecd\mathbb{E}[1/W_{ig}\mid\vecx_i,z_{ig}=1]+\vecLambda_g(\mathbb{E}[(1/W_{ig})\vecV_{ig}\mid \vecx_i,z_{ig}=1]-\veca_{\lambda_g}\mathbb{E}[1/W_{ig}\mid\vecx_i,z_{ig}=1])\},\\
%\end{split}         
%\end{equation*}
%\begin{equation*}
%\begin{split}
\mathbb{E}&[(1/W_{ig})\vecV_{ig}\tilde{\vecU}_{ig}\mid\vecx_i ,z_{ig}=1]
=\mathbb{E}\{\mathbb{E}[(1/W_{ig})\vecV_{ig} \tilde{\vecU}_{ig}\mid\vecx_i,\vecv_{ig},w_{ig},z_{ig}=1]\mid\vecx_i,z_{ig}=1\}\\
&=\mathbb{E}\{(1/W_{ig})\vecV_{ig}[ \vecC\vecd+\vecC\vecLambda_g(\vecV_{ig}-\veca_{\lambda_g})]\mid\vecx_i ,z_{ig}=1\}\\
&=\vecC \{ \vecd \mathbb{E}[(1/W_{ig})\vecV_{ig}\mid\vecx_i,z_{ig}=1]+\vecLambda_g(\mathbb{E}[(1/W_{ig})\vecV_{ig}\vecV_{ig}'\mid \vecx_i,z_{ig}=1]\\&\qquad\qquad\qquad\qquad\qquad\qquad\qquad\qquad\qquad\qquad-
\veca_{\lambda_g}\mathbb{E}[(1/W_{ig})\vecV_{ig}\mid\vecx_i,z_{ig}=1])\},\\
\end{split}         
\end{equation*}
and 
\begin{equation*}
\begin{split}
\mathbb{E}&[(1/W_{ig})\tilde{\vecU}_{ig}\tilde{\vecU}_{ig}'\mid\vecx_i,z_{ig}=1]
=\mathbb{E}\{(1/W_{ig})\mathbb{E}[\tilde{\vecU}_{ig}\tilde{\vecU}_{ig}'\mid\vecx_i,\vecv_{ig},w_{ig},z_{ig}=1]\mid\vecx_i,z_{ig}=1\}\\
&=\mathbb{E}\{(1/W_{ig})(\mathbb{E}[\tilde{\vecU}_{ig}\mid\vecx_i,\vecv_{ig},w_{ig},z_{ig}=1]\mathbb{E}[\tilde{\vecU}_{ig}\mid\vecx_i,\vecv_{ig},w_{ig},z_{ig}=1]'+W_{ig}\vecC)\mid\vecx_i ,z_{ig}=1\}\\
&=\mathbb{E}\{(1/W_{ig})(\mathbb{E}[\tilde{\vecU}_{ig}\mid\vecx_i,\vecv_{ig},w_{ig},z_{ig}=1][\vecC\vecd+\vecC\vecLambda_g(\vecV_{ig}-\veca_{\lambda_g})]')+\vecC\mid\vecx_i,z_{ig}=1 \}\\
&=\{ (\mathbb{E}[(1/W_{ig})\vecV_{ig} \tilde{\vecU}_{ig}\mid\vecx_i,z_{ig}=1]-\veca_{\lambda_g}\mathbb{E}[(1/W_{ig})\tilde{\vecU}_{ig}\mid\vecx_i,z_{ig}=1])\vecLambda_g'\\
&\qquad\qquad\qquad\qquad\qquad\qquad\qquad\qquad+\mathbb{E}[(1/W_{ig})\tilde{\vecU}_{ig}\mid\vecx_i,z_{ig}=1]\vecd'+\mathbf{I}_q \}\vecC.
\end{split}         
\end{equation*}

%\bibliographystyle{chicago}
%\bibliography{mcnicholas}

\end{document}